Article

# Enhancing user experience in large language models through human-centered design: Integrating theoretical insights with an experimental study to meet diverse software learning needs with a single document knowledge base

**Yuchen Wang[1], Yin-Shan Lin[2,*], Ruixin Huang[3], Jinyin Wang[4], Sensen Liu[5]**

[1] School of Architecture, University of Hawaii at Manoa, 2410 Campus Road, Honolulu, HI 96822, United States
[2] Khoury College of Computer Science, Northeastern University, 360 Huntington Ave, Boston, MA 02115, United States
[3] Department of Electrical and Computer Engineering, Carnegie Mellon University, 5000 Forbes Ave, Pittsburgh, PA 15213, United States
[4] Department of Computer Science, Stony Brook University, 100 Nicolls Rd, Stony Brook, NY 11794, United States
[5] Department of Electrical and Systems Engineering, Washington University in St. Louis, 1 Brookings Dr, St. Louis, MO 63130, United States
*** Corresponding author:** Yin-Shan Lin, lilianlin003@163.com

**Abstract:** This paper begins with a theoretical exploration of the rise of large language models (LLMs) in Human-Computer Interaction (HCI), their impact on user experience (HX) and related challenges. It then discusses the benefits of Human-Centered Design (HCD) principles and the possibility of their application within LLMs, subsequently deriving six specific HCD guidelines for LLMs. Following this, a preliminary experiment is presented as an example to demonstrate how HCD principles can be employed to enhance user experience within GPT by using a single document input to GPT's Knowledge base as new knowledge resource to control the interactions between GPT and users, aiming to meet the diverse needs of hypothetical software learners as much as possible. The experimental results demonstrate the effect of different elements' forms and organizational methods in the document, as well as GPT's relevant configurations, on the interaction effectiveness between GPT and software learners. A series of trials are conducted to explore better methods to realize text and image displaying, and jump action. Two template documents are compared in the aspects of the performances of the four interaction modes. Through continuous optimization, an improved version of the document was obtained to serve as a template for future use and research.

## 1. Introduction

Since the emergence of Large Language Models (LLMs) and blowout development from 2022, their integration with Human-computer interaction (HCI) marks the beginning of a new chapter in this interaction. This shift heralds a shift away from traditional HCI, which primarily relied on graphical user interfaces and command-line inputs, toward more sophisticated AI-driven interfaces and models. As Gokul [1] points out, LLMs are reshaping the Artificial Intelligence (AI) landscape with their advanced capabilities in processing and generating human-like language. Their applications extend into various creative domains, including music, art, and storytelling. However, in the aspect of user experience (UX), the LLMs and their applications still present challenges.

## 2. The impact of LLMs on HCI and UX

### 2.1. Redefining UX with LLMs in HCI

LLMs have been pivotal in transforming UX.

One of the big transformations brought about by LLMs is personalization. These models analyze user data and learn from individual interaction patterns to tailor responses and suggestions.

The incorporation of advanced Natural Language Processing (NLP) capabilities in LLMs marks another stride forward. This development allows for more intuitive and human-like interactions.

Additionally, context-aware interactions signify a significant advancement in HCI, brought about by LLMs [2]. These models not only recognize words but also comprehend the context of user requests [3] and predict user's preference [4].

### 2.2. Challenges from UX

As we transition from the exploration of the positive advancements of LLMs in HCI, it becomes imperative to critically examine the multifaceted challenges that accompany this technological integration.

#### 2.2.1. Ethical consideration

As for ethical consideration, while LLMs offer immense potential in HCI, they introduce complex ethical challenges that significantly impact user experience. Ethical challenges mainly come from two aspects: The technology inherent defects, such as specification gaming and side effects [5], pressure to deploy unsafe systems [6] and risks from advanced misaligned AI [7], and inappropriate use, such as Misinformation Harms and Malicious Uses [8].

#### 2.2.2. Supportiveness of user needs

The integration of LLMs into HCI presents a range of technical complexities to meet user's advanced needs, such as the need for higher-speed content generation and more accuracy to the background context, which involves transformers, tokens, reinforcement learning from human feedback (RLHF) and natural language processing (NLP) [9].

Additionally, models often produce outright fabrications that may appear plausible [10]. It's widely acknowledged, through both research and anecdotal evidence, that LLMs often face a pervasive problem of hallucination, or "hallucinated" content.

Subramonyam et al. [11] focus on integrating user experience and needs into the AI development process, finding the problems such as low-level design and share information across expertise boundaries. Zhang et al. [12] uses LLM to answer student questions classified into four types, and finds the system effectively ignores questions that it cannot address.

## 3. Introducing HCD to LLMs and their applications

### 3.1. Principles of HCD

HCD, or HCAI, introduced by Don [13], is a problem-solving approach with its core positioning real individuals at the center of the development process. This approach is focused on delivering equitable results and upholding the utmost respect for privacy, thereby aligning AI functionalities with human values and ethics [14]. The essence of HCD lies in consistently prioritizing the user's desires, challenges, and preferences throughout every stage of the design and development process [15].

Major principles of UCD includes early and active involvement of the user during the design process, clarification of user, user feedback is incorporated into the product's lifecycle and the product is improved using an iterative design process [16]. For instance, Jaimes et al. [17] emphasizes the importance of mixed-initiative human-computer systems, highlighting how user input plays a crucial role in shaping the functionality and responsiveness of these systems. Similarly, Mack et al. discusses the criticality of including diverse user perspectives in research methods, ensuring that systems are accessible and meet varied user needs [18].

Some research explores the way to encourage public early participation in public decision making or affecting users' climate-controlling behavior by using new technology, such as augmented reality (AR) [19] and virtual reality (VR) [20], which are also applications of HCAI. Research by Seffah and Andreevskaia [21] developed a skill-oriented program towards developers and students based on analyzing UCD knowledge and techniques.

### 3.2. Previous attempts to reflect HCD in LLMs applications

This study mainly focusses on the methods of enhancing the supportiveness of user needs by applying HCD principles. HCD prioritizes the needs, preferences, and contexts of users [22], ensuring that LLM-driven interactions are not only efficient but also resonate with the users' expectations.

Petridis et al. [23] explore the possibility of incorporating prompt-based prototyping into designing functional user interface (UI) mock- ups, finding LLMs potentially reduce the time needed to create a functional prototype. Park and Choi [24] introduce LLMs into audience simulation for public speech and uses AudiLens to provide flexibility to the speaker. Di Fede et al. [25] introduces the Idea Machine combined with LLMs to empower people engaged in idea generation tasks.

Korbak et al. [26] explores alternative objectives for pretraining LMs (Language Model) to create text aligned with human preferences. Study by Rastogi et al. [27] finds existing auditing tools use either or both humans and AI to find failures. They create the evaluation tool: AdaTest++, which is powered by GPT3 and Azure's sentiment analysis model.

### 3.3. Build HCD guidelines to enhance UX in LLMs

From the above discussion, the HCD principles related to LLMs can be concluded as the following (**Table 1**).

These six guidelines are crucial to LLMs like GPT in meeting user expectations and needs effectively. They also provide possible ways to optimize related design including AI agent, application, platform, user interface and the construction of knowledge base.

**Table 1.** HCD guidelines related to LLMs for enhancing UX.

| Principles | Requirements |
|---|---|
| High efficiency | Fast response |
| Feedback consideration | Collect feedback; Update periodically or imperiodically |
| High supportiveness for diverse needs | Generality and specificity consideration; Personalization and Customization |
| Emotional consideration | Understanding the emotional of user conversations and providing more humanized interactions |
| High Simplicity | Easy input; Effortless expression; Multimodal input and output: images, text, voice, etc. |
| High Reliability | Authenticity; Accuracy |

In the following section, a preliminary experiment is conducted to apply these HCD principles into the enhancement of LLM's interaction capabilities.

## 4. Improving UX by optimizing a single document as principal knowledge in GPT: A preliminary experimental study

This preliminary experiment mainly focuses on the UX enhancement from the aspect of supportiveness for users' diverse needs. Other HCD principles, such as simplicity and reliability, will also be taken into consideration in the experiment design. The experiment takes ChatGPT-4 as an example, exploring how to use a single document as the main material of knowledge base to construct a custom GPT.

### 4.1. Materials and methods

#### 4.1.1. Virtual experimental environment: ChatGPT-4 and GPTs editor

The working environment for this study is set in ChatGPT-4 and GPTs Editor.

GPTs editor is a relatively new function as one part of ChatGPT-4. It's a specialized environment for creating and tuning GPT models based on GPT editor's preset configuration, including descriptions of this GPT, instructions, knowledge, starters and actions, allowing adjustments to the model's responses, capabilities, and interaction style. In "Configure" interface, the "Instructions" area provides overall control rules for GPT to follow during interaction. "Conversation starters" allows users to start a conversation by just clicking corresponding buttons. "Knowledge" provides a preset knowledge base where editor can upload files as data in certain formats, such as docx, pdf or jpg.

After the new GPT being created, it will be imported in ChatGPT-4 automatically, which provides an environment for users to interact.

#### 4.1.2. Principal objective: Meeting users' diverse needs in software learning interaction

This study defines a goal as taking ChatGPT-4 as a software learning tool that provides knowledge and solutions for novices in learning a new software. This hypothetical scenario is designed to simulate how LLMs can synthesize newly inputted knowledge and utilize it in multiple ways, which can be considered as one of the typical applications which use LLMs to serve a specific group of people. UX in this study can be evaluated by the quality of dialogues during interaction.

Visual Scripting, a tool inside Unity software, is taken as the software in the optimization process. It allows for the creation of logic and game behaviors without writing code directly. By using visual graphical nodes and connecting them with lines, Unity developers can construct complex game logic and interactions. The advantages are as follows:

(1) GPT has less inherent knowledge about Visual Scripting itself, even if some related coding knowledge is trained into GPT. Therefore, the pre-existing knowledge will less interfere the evaluation.

(2) The Visual Scripting Manual on official website can be used as a reliable source for constructing the knowledge for GPT.

(3) The images of how to use Visual Scripting are easier to make and since it is an intuitive tool.

#### 4.1.3. New knowledge resource: A single document input in knowledge base

A single document is used as the main new knowledge resource uploaded in GPTs Editor's "Knowledge" area. It is a Microsoft Word document in docx format, serving as the new knowledge resource and control module. It is composed of a control part and a software knowledge part. The advantages of using a single document are as follows:

(1) Simplicity and customization consideration: It is easy for a real creator to replace certain parts of the template document to make another GPT as a tutorial for learning other software.

(2) Compatibility consideration: The docx document can contain the knowledge both in forms of natural languages or codes. The arrangement of content is also easy to be adjusted.

(3) Variables control: To avoid black box effect which often exists in AI product, the single document can be easily optimized, which helps to explore a method of getting a relatively controllable result.

#### 4.1.4. Users' needs and requirements definition

Different groups of users may have different needs for the usage of a software learning GPT, while a single user may also have needs for multiple ways to use it. The following diagram shows the possible needs (**Figure 1**).

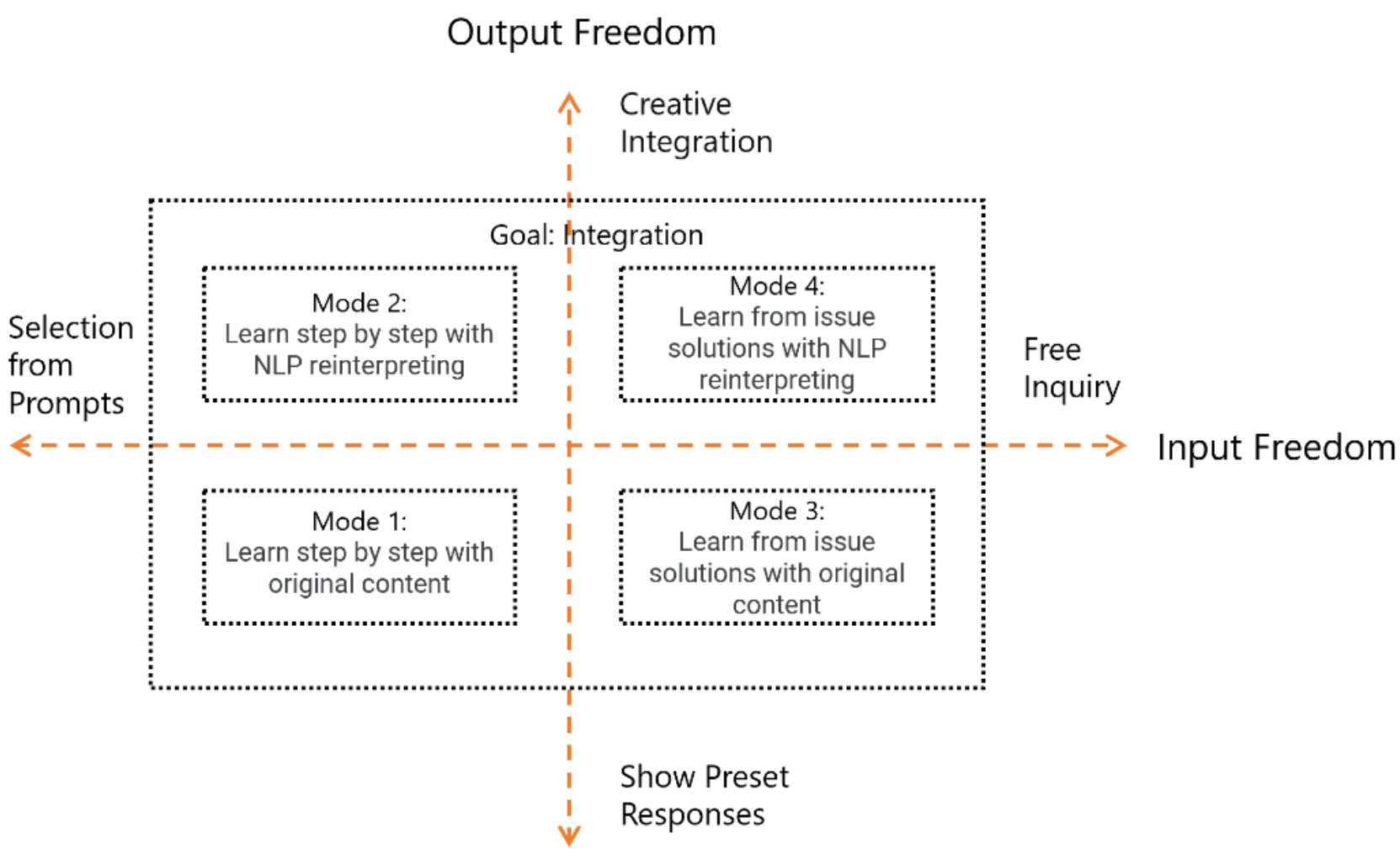

**Figure 1.** Software learners' diverse needs in using this GPT.

It can be considered that different using modes of this GPT are based on users'

requirements for varying degrees of input and output freedom. The overall goal is to integrate these modes.

For the input freedom, inputting single a number or letter based on the given prompts is an alternative to select a desired action, such as to start the tutorial, or jump to a certain section of the tutorial, which requires less input freedom. Users also have the demands for inputting a complex issue and then getting solutions, which requires more input freedom.

For the output freedom, the alternative of strictly showing the original content from the knowledge part of the document is needed, meaning less output freedom, which can be applied in the scene that users hope to strictly obey the software guidance from a traceable source. In other cases, the output needs to display content in a creative way by using more natural and coherent language to rewrite and reorganize the knowledge, meaning more output freedom.

Therefore, four types of modes are supposed to be realized:

Mode 1: Learning step-by-step with original content, enabling users to learn from printed original content retrieved from the knowledge bases words by words.

Mode 2: Learning step-by-step, similar to the previous one, but use NLP to reinterpret original content.

Mode 3: Learning by issue solutions, allowing users to receive solutions for their issues while using Visual Scripting, and the solutions should print the original sentences of related knowledge.

Mode 4: Learning by issue solutions, similar to the previous one, but use NLP to reinterpret original content.

#### 4.1.5. Expected outcomes

The overall optimization process can be illustrated as the following diagram (**Figure 2**).

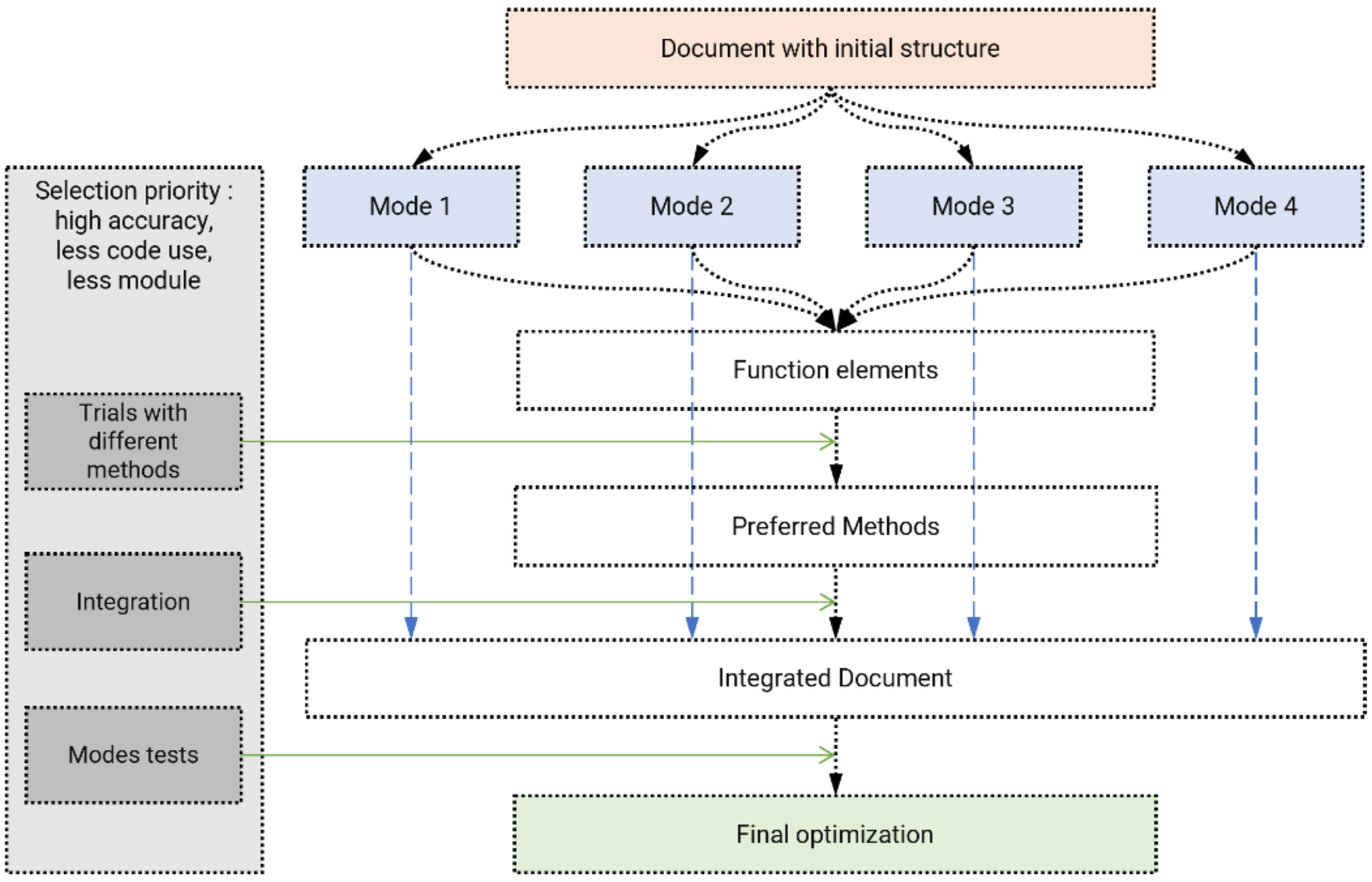

**Figure 2.** Document optimization process.

The process involves a series of examinations. Firstly, different functions to realize these modes will be analyzed. Different methods will be tested to within the document initial structure. Then, these methods will be filtered and selected based on the results of tests. The document will be adjusted by using the preferred methods. The four modes will be tested within the new document to check the result of performance and finally, it will be optimized again based on the result and then be retested. The selection principles in each step include high accuracy, less code use and less module use.

The expected outcomes of this study are as follows:

(1) To figure out how the elements inside the single knowledge document as well as the GPT configurations will affect interaction quality.
(2) To explore the possibility to integrate this GPT's four modes.
(3) If possible, through optimization process, a document that can better integrate these modes get be finally obtained as a template for future use, which can be seen as the application of "feedback consideration" HCD principle.

#### 4.1.6. Methods of UX evaluation

The method of UX evaluation is to assess the accuracy of interaction rather than speed. To emphasize the ability of GPT comprehensively utilizing newly input knowledge, is supposed to control the variables in the input-output process. The following methods are employed to prevent GPT's directly using knowledge of Visual Scripting to interfere the evaluation:

(1) Multiple forms of knowledge composition

The images of Visual Scripting nodes and connections are taken as knowledge together with text. Some of the original text and images from Unity Visual Scripting Manual 1.9.1 version [28] is extracted or rewritten and then be placed into the document.

(2) Closing web browsing

Web browsing action in GPT may introduce original online resources, so closing it can isolate environment.

(3) Methods of output control

The output is supposed to mix text and images from the knowledge, making it challenging to achieve user's goals.

The average accuracy of the results in each trial will be assessed through 5 times of repetitive complete chatting, the functional elements of the four modes and their evaluation criteria of the interaction result are as follows (**Table 2**).

**Table 2.** Criteria for approximate accuracy assessment.

| **Functional Elements and Criteria List for modes** | |
|---|---|
| (A) Whether the jump action is successful and smooth;<br>(B) Whether the output obey the sequences of the original steps;<br>(C) Whether related images can be successfully displayed together with text.<br>(D) Whether the output display original content in each step completely;<br>(E) How much the reinterpreting using NLP deviates the original contents in the document, producing wrong content or "hallucinated" content (content that seems to be correct but has no relevance with the document original content);<br>(F) How helpful the selected content is to the user's question (the designed questions are designed to be satisfied by some certain parts in the document);<br>(G) Whether the output solutions cross enough necessary range of knowledge chapter in the document. | |
| Mode 1 | Mode 2 |
| (A), (B), (C), (D) | (A), (B), (C), (E) |
| Mode 3 | Mode 4 |
| (C), (D), (F), (G) | (C), (E), (F), (G) |

For testing the function in Mode 3 and 4 that answer user's questions, the question list is designed as follows (**Table 3**).

**Table 3.** Question list for Mode 3 and 4.

| Question Number | Content |
|---|---|
| Q1 | How can I use nodes to change the sprite of GameObject "A" when a time duration finishes? |
| Q2 | How can I use nodes to sets the velocity of GameObject "B" to half of its original velocity when a "B" enters a trigger collider in 2D space? |
| Q3 | How can I use nodes to add value 1 to the existing object variable named "C" and set back to "C"? |
| Q4 | How can I make another Script Graph named "D" inside a Script Graph named "E" to receive a float from "Script E", then returns true if the float is greater than 1 and less than 2, otherwise returns false to "Script E"? |
| Q5 | If UI button "G" has a Script Graph named "H" and GameObject "J" has a Script Graph named "K", how can I use nodes in "H" to trigger the event in "K" when clicking the button "G"? |

These questions are with high complexity and less specificity, meaning to require crossing different sections of knowledge part in the document to find answers, and less mention any specific name of node and the phrase "Visual Scripting". The intent is to make it easy to recognize whether it use the new knowledge (**Figure 3**).

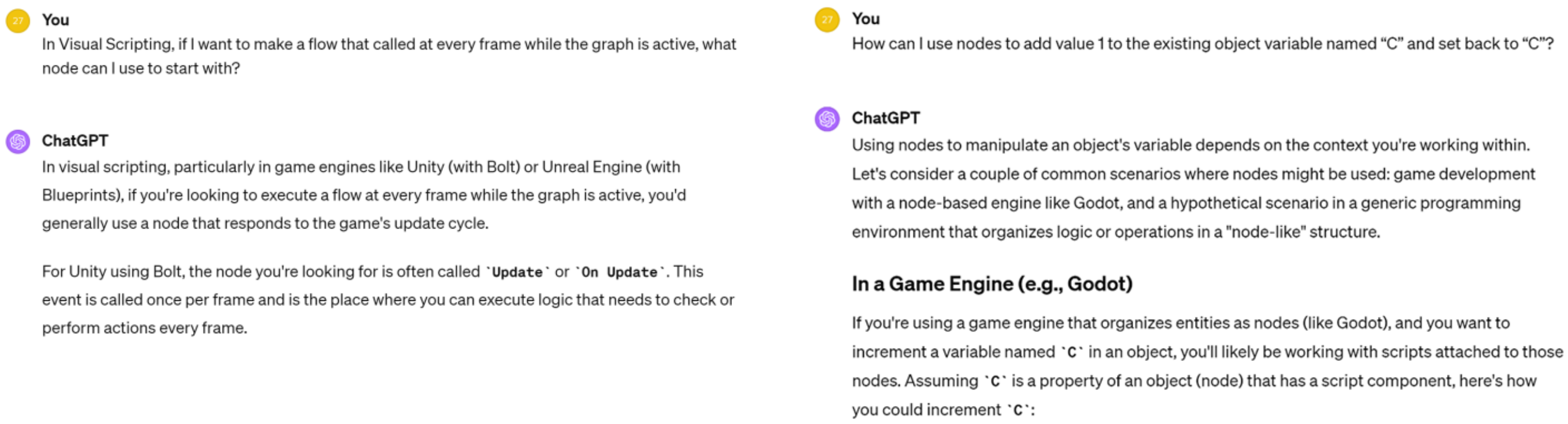

**(a)** **(b)**

**Figure 3.** GPT's responses to different questions without new document input. **(a)** Question with less complexity and high specificity; **(b)** Question with high complexity and less specificity.

#### 4.1.7. Structural design of the single document

The initial structure of the new document is designed as follows **Figure 4**.

The document includes two parts: Part 0 provides an introduction and response methods, serving as a general control part; Part 1 provides the knowledge of Visual Scripting, structured into chapters and sections based on the content. After each output, users can change interaction mode directly.

Here is a brief introduction of these parts:

Part 0 includes:

(1) "Overview of This GPT's Rules" section: It provides general rules to control the interaction. It includes the following parts:

a) Descriptions of this GPT and the document.

b) The general rules, such as "Refer to Section 2 in Part 0 for initial dialogue rules" and "Interactive requests needing user responses are enclosed in braces {}, such as {Enter 1: Continue}". Prohibited interaction ways area also provided, such as "Refuse interactions that does not meet current interactive requests inside {}".

(2) "Initial Response Method for Dialogues" section: It provides ways to go to different parts corresponding to the input Starters.

(3) "Start learning" section: It provides ways to process Part 1 section by section.

(4) "Finding Solution" section: It provides ways to provide solutions to user's questions.

(5) "Introduction" section: It provides the basic information about how to use this GPT.

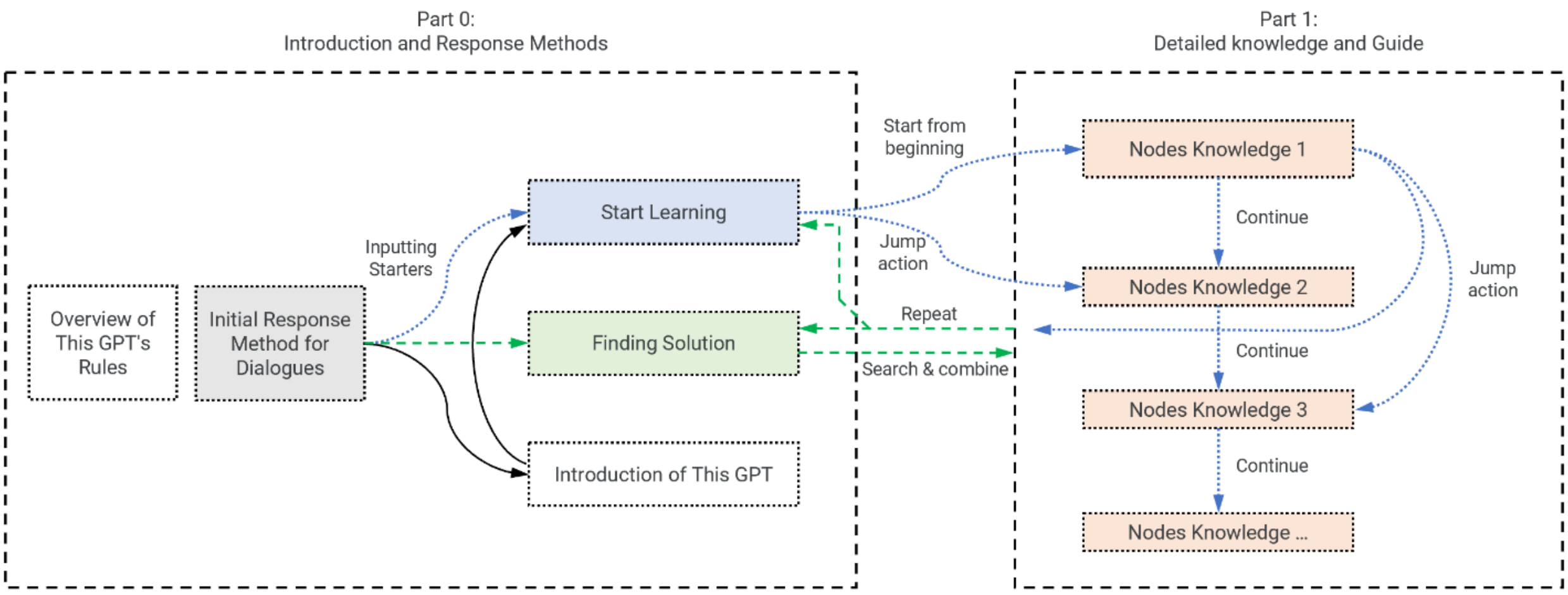

**Figure 4.** Initial structure of the single document.

Part 1 includes:

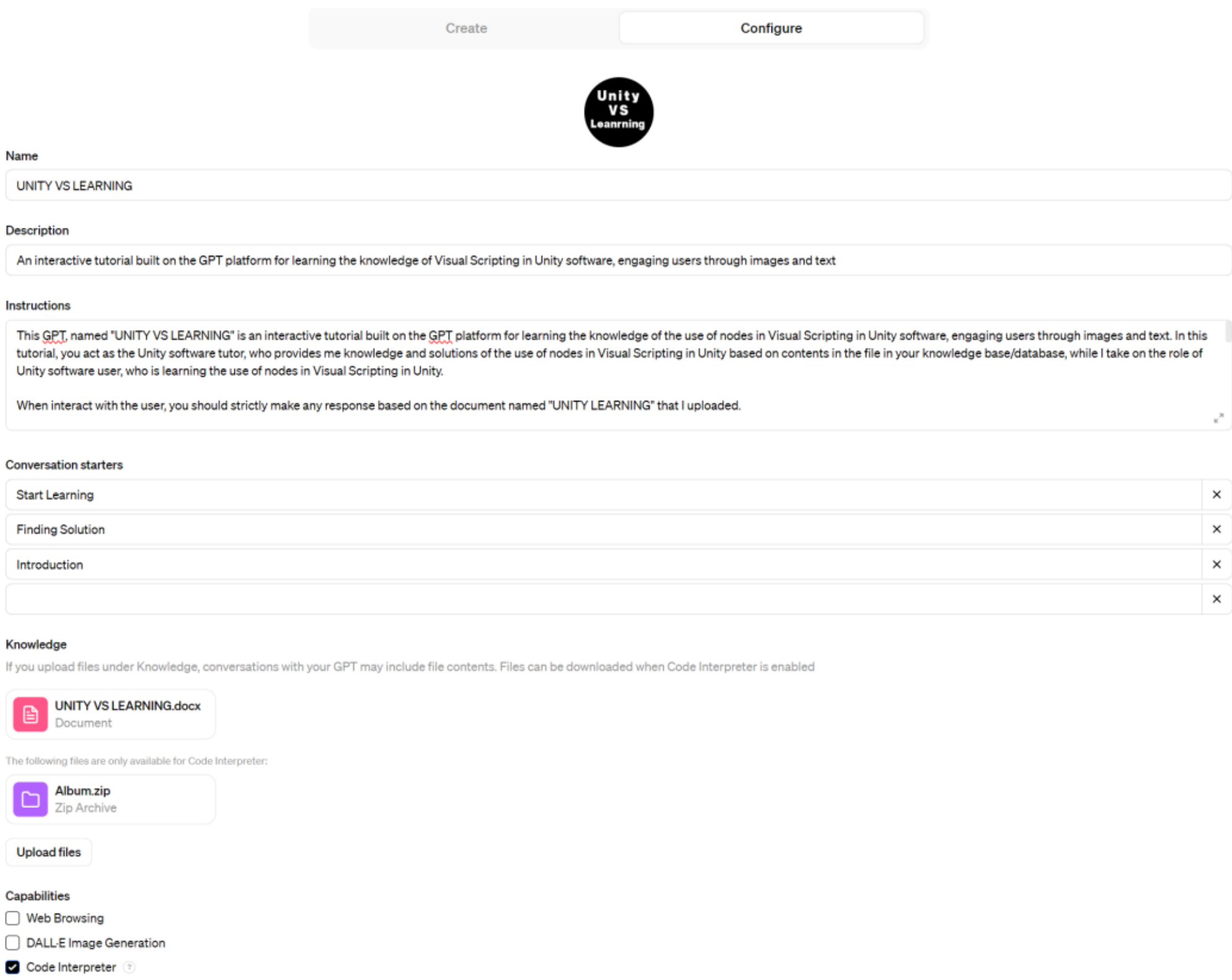

**Figure 5.** Configuration of GPT.

14 Chapters of detailed knowledge are provided as well as some guidance. There is a title of each chapter and each section, but no overall introduction in each Chapter.

It is found that the extraction of images directly from the Word file is not possible. Therefore, a zip file containing multiple images is uploaded as supplementary material. Each image is named in following format: "Chapter number" + "Section number" + the image's sequence number, such as "060401". Also, the "Instructions" area is filled with content from "Overview of This GPT's Rules" section. In addition, the code interpreter is turned on for processing code in the document. The configuration of the new GPT is set as **Figure 5**.

## 4.2. Module and structure trials: Results and analysis

### 4.2.1. Original text display

For displaying the original text in the document, different ways are examined by several trials.

(1) Trial 1: Executing the "print ( )" function when jumping from somewhere else

Here is an example. The instructions in Part 0's "Overview of This GPT's Rules" reads: "Text that needs to be directly printed will be with clear instructions such as 'print (Hello)', and will be enclosed in brackets marks ()." The instruction in "Initial Response Method for Dialogues" in Part 0 reads: "If 'Start Learning' is inputted by the user, go to Part0, Section 3A." and in Section 3A, it uses "print ( )" function in each step following a serial number, such as "1. Hi. Welcome…" and "2. The following is…". The result shows it can proceed printing text step by step easily as following screenshot (**Figure 6**). The limitation is it has to proceed from the first step in a section.

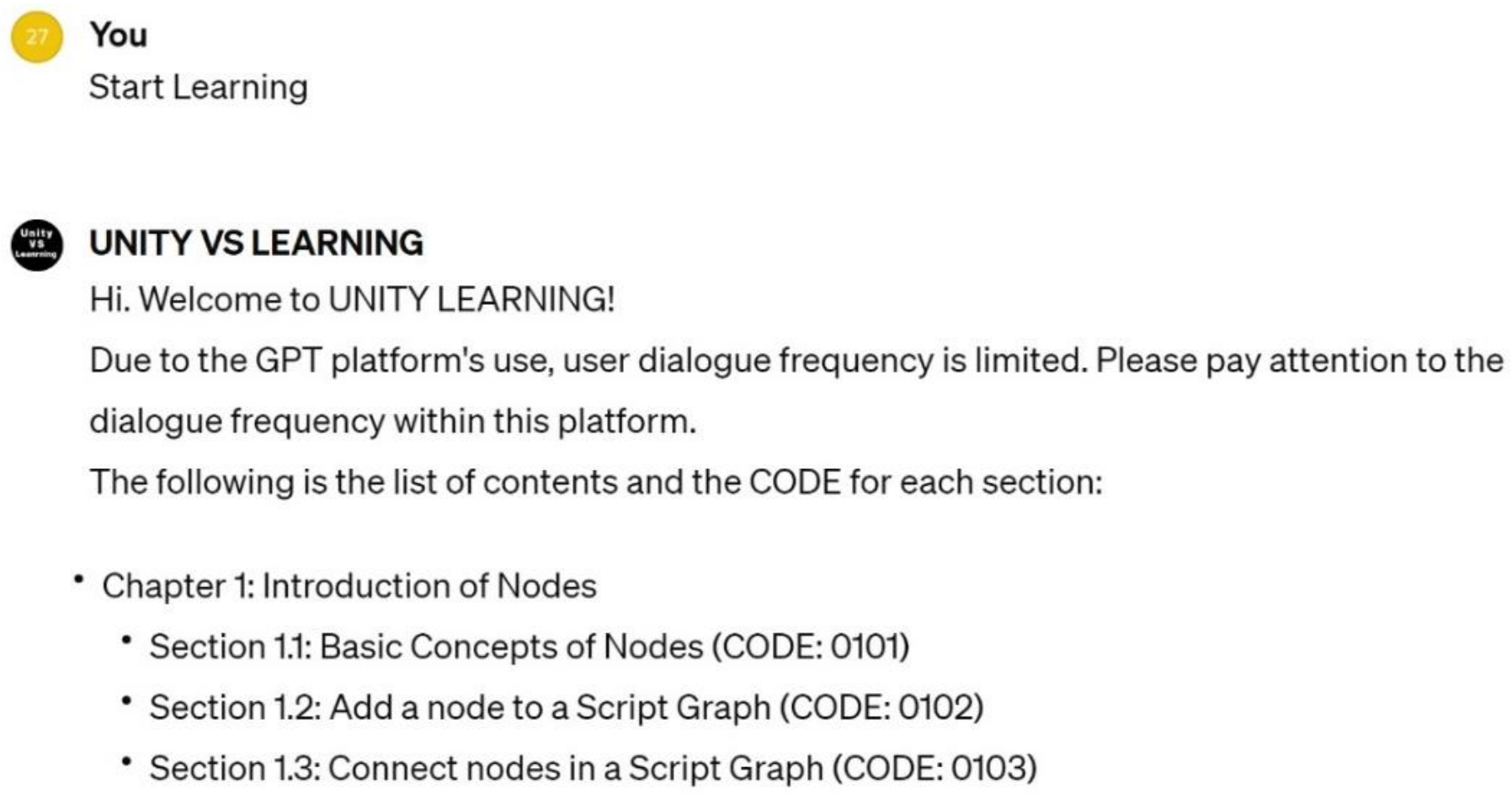

**Figure 6.** Result of text display Trial 1 (part of the whole result).

It is also examined whether it can work if removing the "print ( )" function in Trial 1. It shows that when jumping from somewhere else, this approach cannot always keep the text printed in its original form.

(2) Trial 2: Using printing command to print text in another place

Here is an example. The instruction in "Initial Response Method for Dialogues" in Part 0 reads: "If 'Start Learning' is inputted by user, print Point 4 in Part 1, Chapter 1, Section 1.1, then print Point 4 in Part 1, Chapter 1, Section 1.3" or "If 'Start

Learning' is inputted by user, proceed the following steps: Step 1. print Point 4 in Part 1, Chapter 1, Section 1.1; Step 2. print Point 4 in Part 1, Chapter 1, Section 1.3."

Some variables here include: a) whether directly providing the text content in Section 3A, or with "printing ( )" function; b) whether use single number at the beginning of the text like "2.", or use serial number like "Step 2", "Point 2"

All results with different combination of variables successfully printed the text, however, they have to continue printing until the end of the section. Also, if it is required to print sperate parts in Part 1, it can just finish the first one.

(3) Trial 3: Using command to execute the "print ( )" function in somewhere else

Here is an example. "If 'Start Learning' is inputted by user, proceed Step 1 in Part 1, Chapter 3, Section 3.1" and in the corresponding section, it use "print ( )" function.

With different variable forms, the results are similar to those in trial 2. However, it can display two separated parts of text in one time.

(4) Trial 4: Searching text to display

An approach is to use instructions in Part 0 to force GPT answer user's question with original text content as follows: "If users ask you any question, please print any useful information in Part 1 that can answer user's question. Please print the original text and do not rewrite them or add your own words. Please notice that the useful information in Part 1 can be over one place, so please find as much as possible."

The results are as follows (**Figure 7**). The red square is to mark the original text. It can be seen that even under strong instructions, it still tries to rewrite the original text to make the content coherent. The reason might be it has a strong weight of using NPL since it is a LLMs.

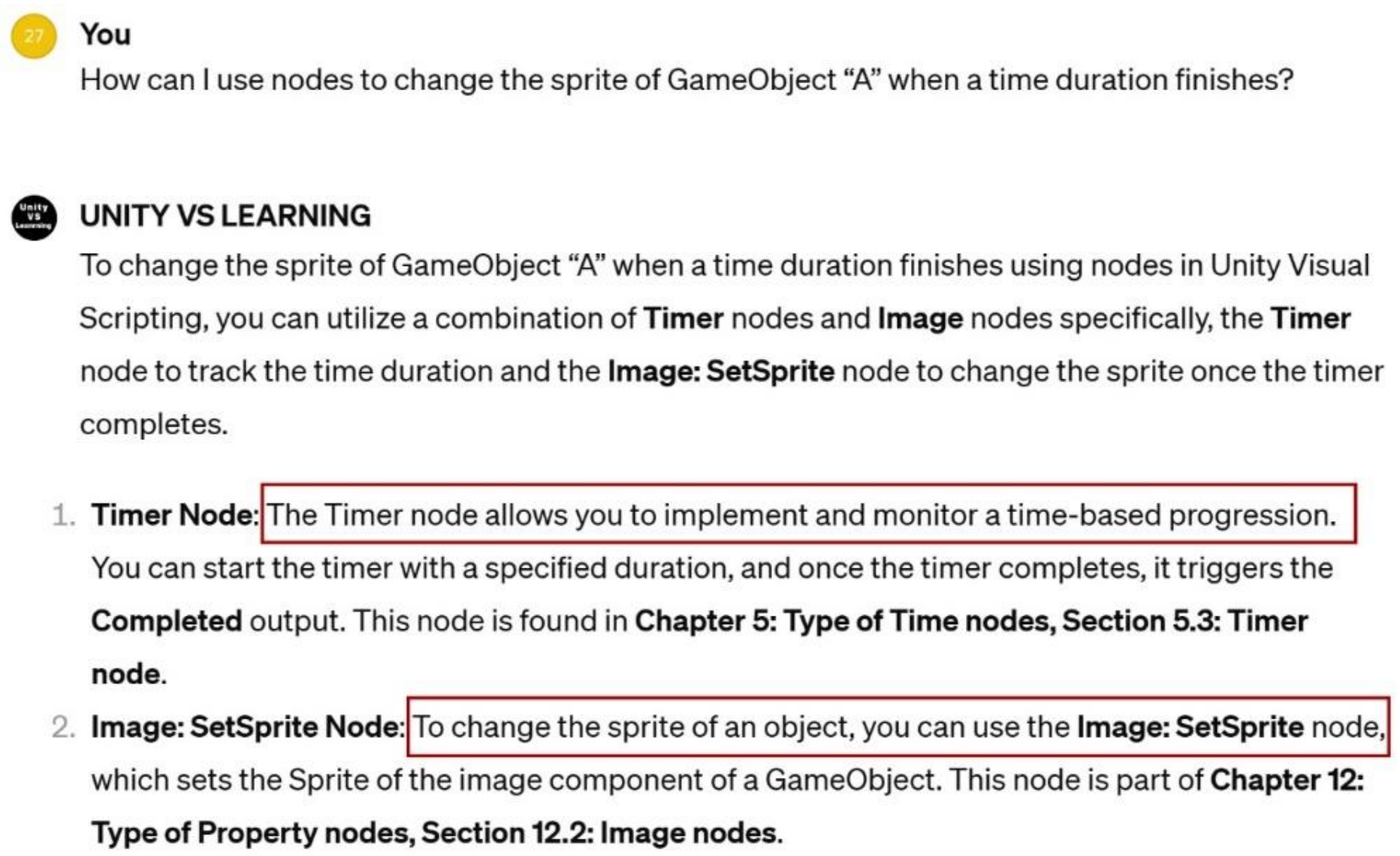

**Figure 7.** Result of text display Trial 4 with easy input (part of the whole result).

However, if the requirement of using original text is added into user's input, GPT will largely increase the weight of using original text, as shown in **Figure 8**, which shows that the user's input plays a decisive role during interaction.

Another test is to use an existing printed instruction as context to force GPT print the original text. It induces GPT to print out the command first as context and then the user adds a signal from the printed instruction in new input. The result shows this method also does not work.

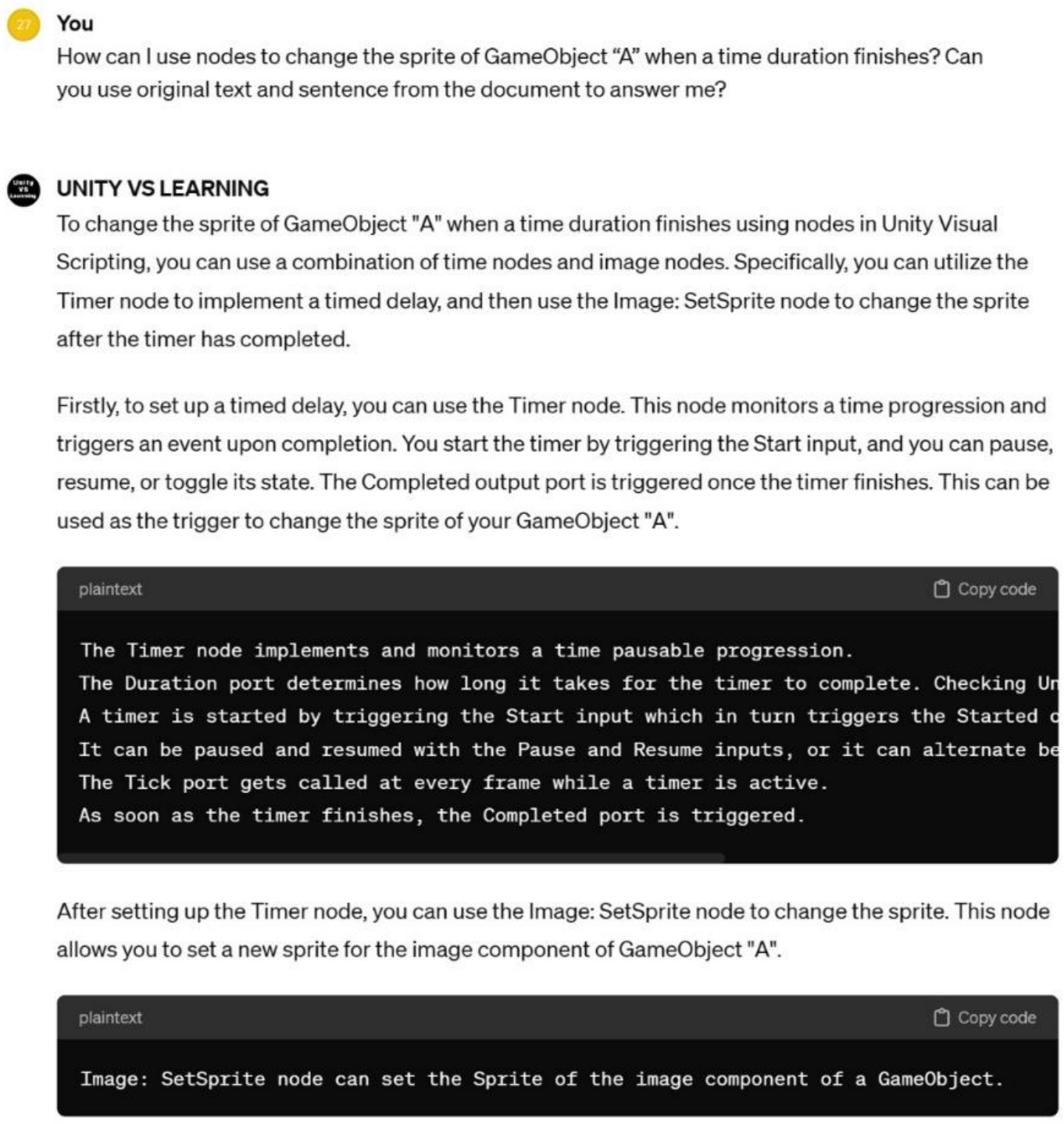

**Figure 8.** Result of text display Trial 4 with strong input (part of the whole result).

It can be inferred that without advanced code, it is difficult to force GPT print the original text when asked a question that is not preset, because this process may involve several steps with different ability. Therefore, it is preferred to design the Mode 3 to work with user's additional input instruction. The results also show that "print ( )" function of every step in Part 1 rarely interfere the information searching process.

Move the conditions into "Instructions" area are also testes and the results show they sometime works.

(5) Trial 5: Reinterpreting text in natural language

Reinterpreting text in natural language is required by Mode 2 and Mode 4. Due to GPT's NLP characteristic, if there is no additional command, it is easy to realize the reinterpreting function. However, if there is command like "print ( )" in the content, or instruction to print original text, it is needed to add a conversion mechanism.Considering the convenience for future editors of this document, it is not supposed to provide conditional statements with many options after each section in Part 1 for switching modes. It is found that when conditional statements are only placed in part 0, it is difficult to reinterpret the text in Mode 4. It works only if the user add prompt like "in natural language" into the input. However, when trying to move conditional statements into "Instructions" area, it works well.

### 4.2.2. Image display

Images are compressed into zip file named "Album" uploaded additionally. Four ways to display images are examined here.

Trial 1: Directly commanding to show images

When using command like "show the image named '010101.jpg' from the zip file", "display image named '010101' from '/mnt/data/Album.zip'" or "extract the image named '010101.jpg' from the zip file "Album.zip" in "/mnt/data/" and display", it is sometimes failed and no image is shown.

Trial 2: Using half code to extract every time

When using sentence like "Please use code including 'display.display(image)' to display image named "010101.jpg" in path '/mnt/data/Album.zip'" to display code, it sometimes works, while in other time it fails. Sometimes, even if it generates the complete required information, the sequence of image and text is disordered and it always display an image at first. This may be because GPT takes this order as a natural-language description rather than a code, so it mixes the information with other to utilize.

Trial 3: Using prompt to call a shared piece of code

A shared piece of code to display images lis put in the Part 0 (**Figure 9a),** for example, in Section 7. And in Part 1, it provides the following prompt in each place where needed: "Please use code in Part 0, Section 7 to display an image (image_name = '010101.jpg')." This approach can show the image successfully. However, it also has the disadvantages of disorder, giving extra description (marked with red square in the figure) (**Figure 9b**) and sometimes turn into reinterpreting in the middle.

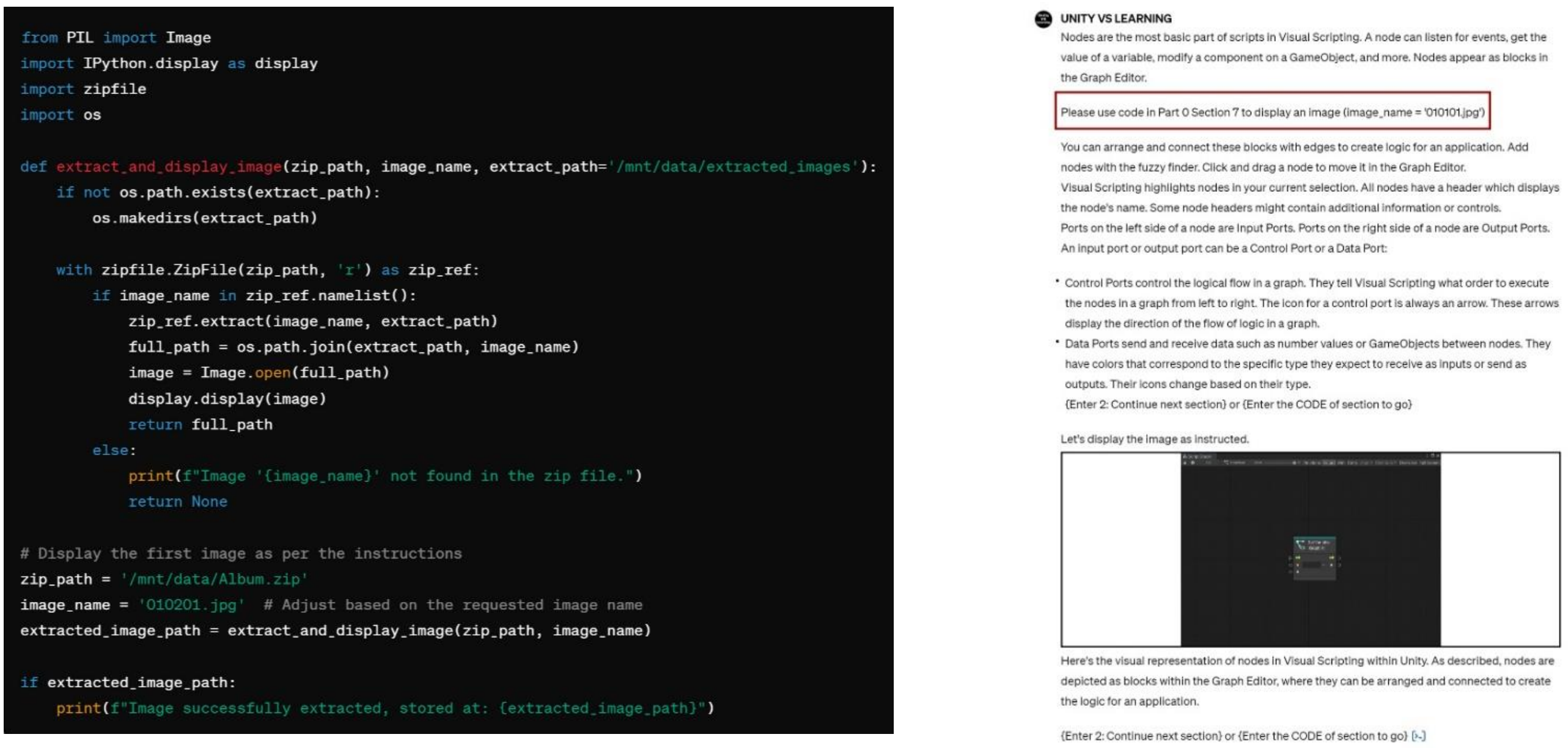

**(a)** **(b)**

**Figure 9.** Code and Result of image display Trial 3. **(a)** Code of Trial 30; **(b)** Result of Trial 30.

Trial 4: Using complete code to extract one image every time

This approach is using code in (**Figure 9a**) in each place where needed. The code is executed every time when being proceed together with text printing. In most results, the image can be displayed in correct sequence. The disadvantage is that the document needs to repetitively provide the code.

Trial 5: Extracting all images from the beginning

This is to use a complete code (**Figure 10a**) to extract all images from the beginning, then use following request to show the extracted images where needed:

“Please use code <image_path = ‘/mnt/data/extracted_images/010101.jpg’

display.display(Image.open(image_path))> to display an image in path ‘/mnt/data/extracted_images’.”

The performance is relatively good with correct sequence and makes less errors (**Figure 10b**).

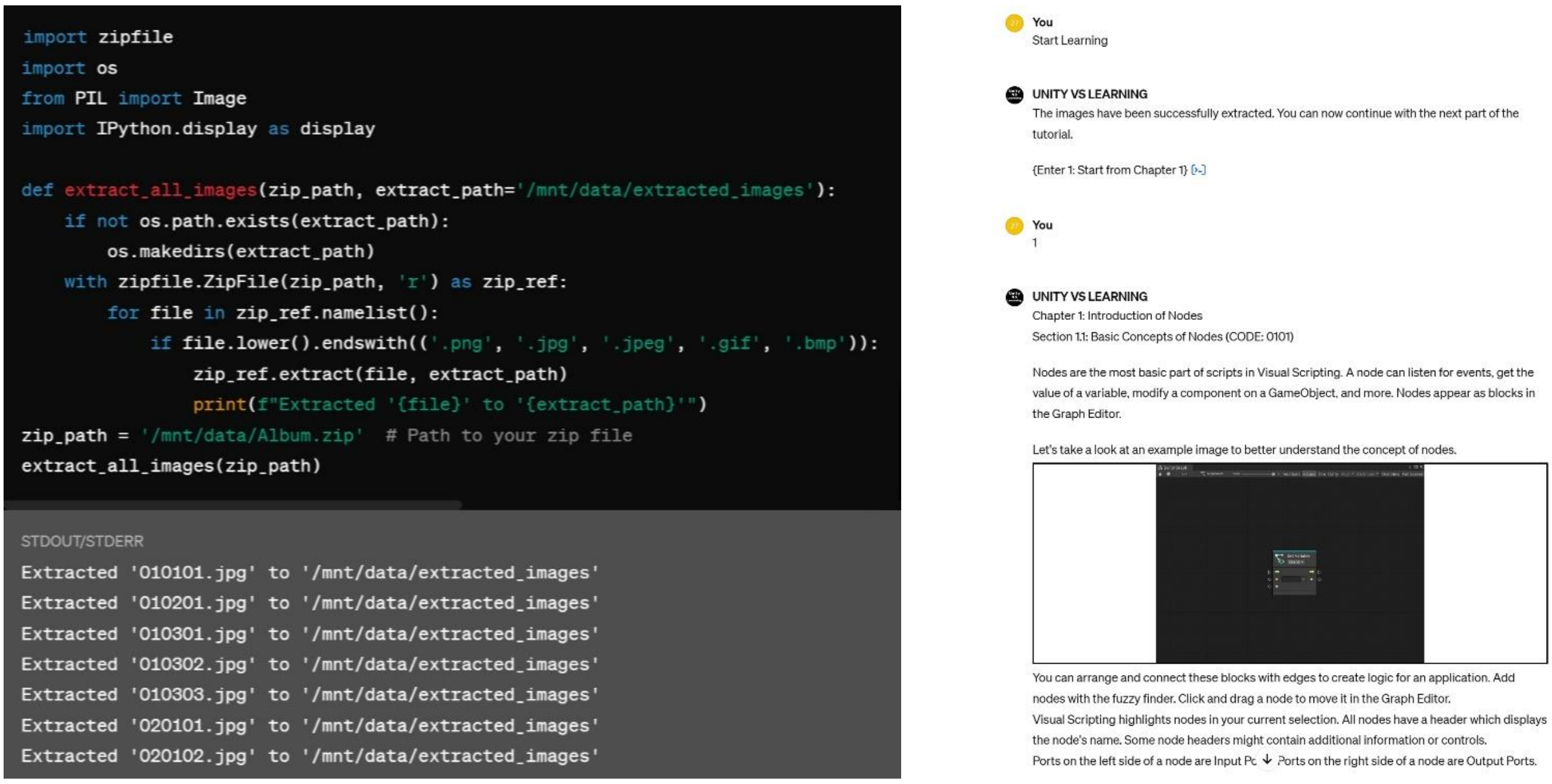

**(a)** **(b)**

**Figure 10.** Code and Result of image display Trial 5. **(a)** Code of Trial 5; **(b)** Result of Trial 5.

Trial 6: Display image by searching

It is also required to display an image when giving solutions to user’s question. Two ways are explored based on Trial 5’s method which extracts all images first.

One approach is to add additional instruction to all of the conditional statements in “Control Center” section (will be discussed later), “Overall Rules” section and “Instructions” area as follows:

“Please also execute the steps with code for displaying images that is very close to the information you find and in the same section of the information you find, which helps to illustrate the text information.”

The results show it works in printing the text it found and displaying the corresponding images. **Figure 11** shows two pieces of one result. There is one disadvantage that it sometimes put all images together, even when an addition instruction.

Another approach is similar to the previous one, but placing code in part 0 and in Part 1, telling GPT to execute this code to display an image. The results and disadvantages are almost the same as the first approach.

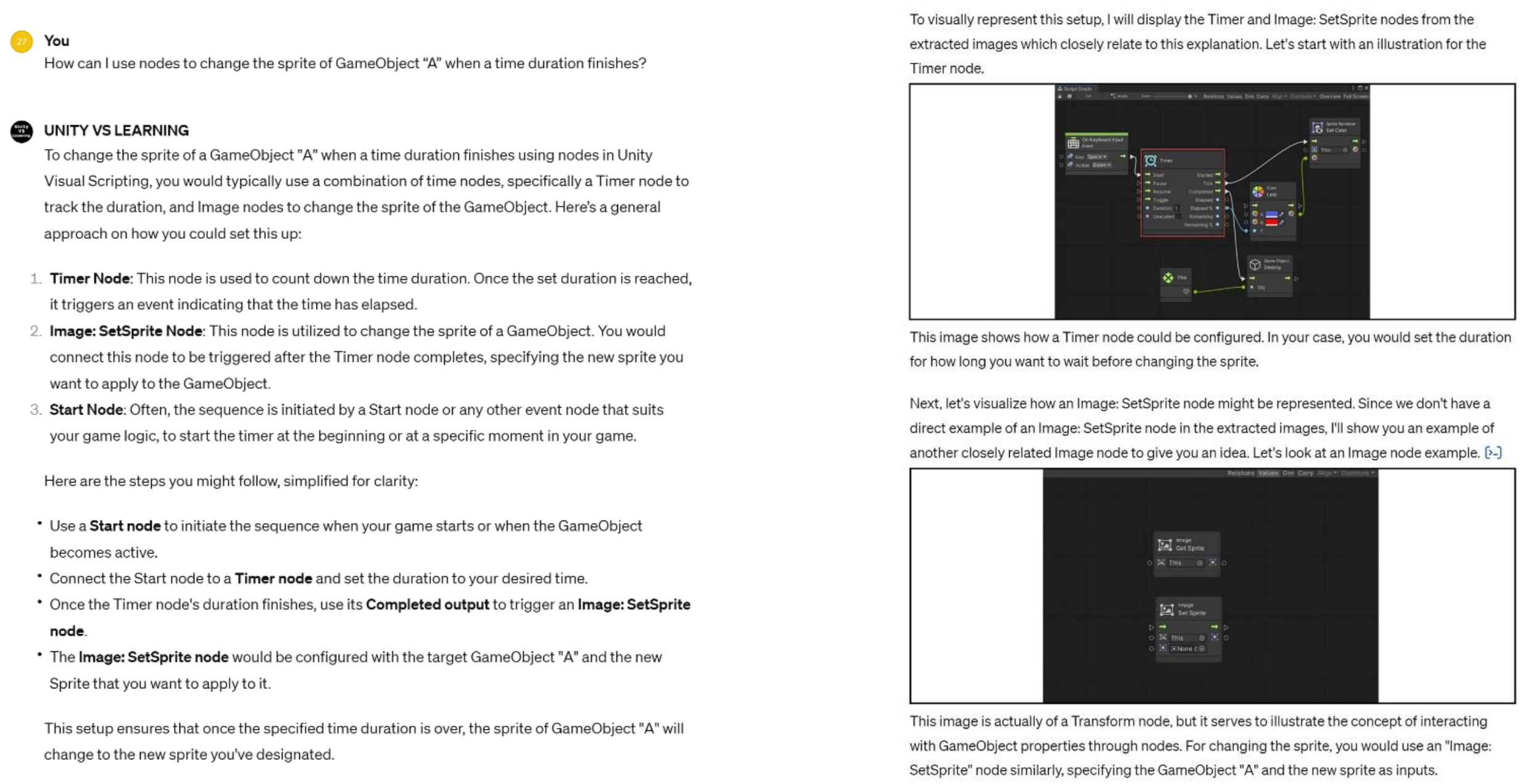

**(a)** Piece 1. **(b)** Piece 2.

**Figure 11.** Results of image display Trial 6.

### 4.2.3. Jump action

This part explores how to jump from one place to another in different ways. Using conditional statements is to add an additional part where needed with several "if" conditions corresponding to users' input, and they just exist in the document and are not be printed. Using interactive request is to provide a request enclosed in braces such as {Enter 1: go to Section 3} at the end of each section. Using section title is to place prompt in section title for positioning.

(1) Trial 1: About section title and "Initial Response Method for Dialogues"

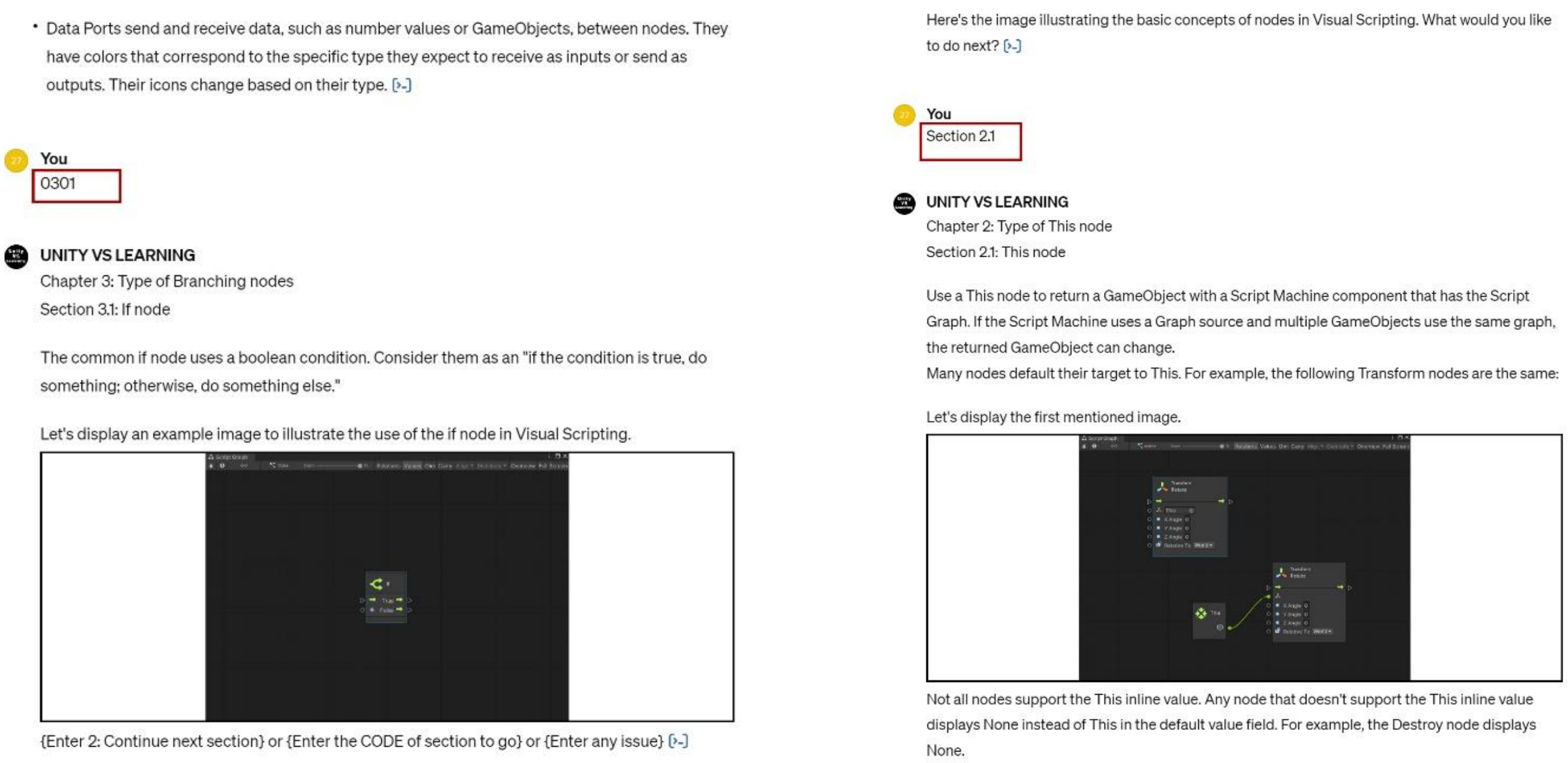

**(a)** **(b)**

**Figure 12.** Result of Jumping to a specific section. **(a)** Piece 1; **(b)** Piece 2.

The weight of section title and "Initial Response Method for Dialogues" section designed previously are tested. It is found that actually sometimes GPT tends to go to a section with title that is same as input words, more than go to a section according to conditional statement in "Initial Response Method for Dialogues" section. Therefore, it is better to use a different name in section title if conditional statements are supposed to work.

Jumping from the end of a section to any specific section in Part 1 is also a required function for user's step-by-step learning. In a test, each Section in Part 1 is labeled with a unique code at the end, such as "0201" meaning the Section 1 in Chapter 2. It is found the jump action can work no matter whether there are any conditional statements or an interactive request like "{Enter the CODE of section to go}" (**Figure 12a**). When the section title is input, it also works (**Figure 12b**). It can be inferred that GPT actually jump to corresponding section by searching section title.Trial 2: About Interactive requests

The "Instructions" area in "Configure" interface is filled with the rule that the user can only interact with interactive requests. When interactive requests and Control Center provides different directions, it is found that the interactive requests have a large weight when it is explicitly stated, such as {Enter 1: Continue} (**Figure 13a**). However, if it is obscure, GPT will locate user's input to other conditions (**Figure 13b**).

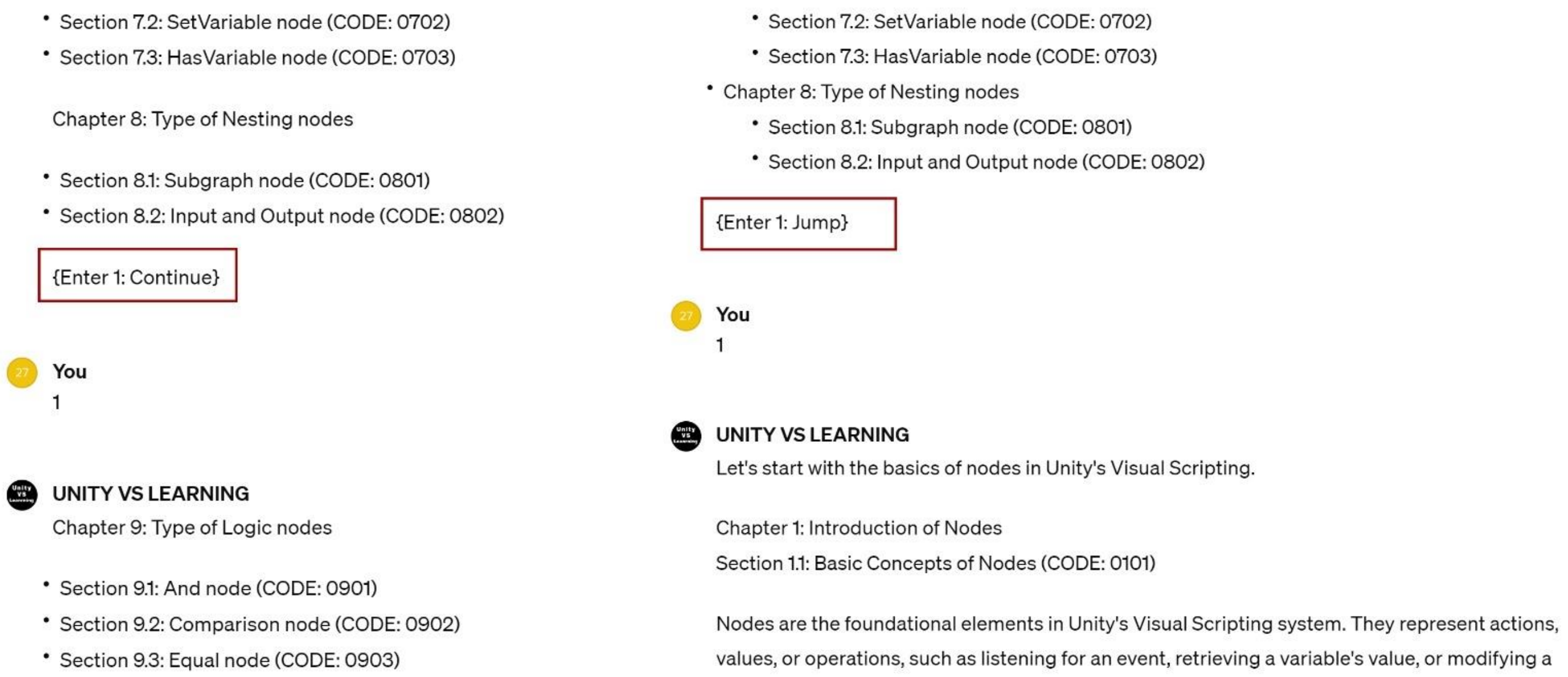

**(a)** **(b)**

**Figure 13.** Results of different interactive requests in Trial 2. **(a)** Trial 2a; **(b)** Trial 2b.

Trial 3: About positions of conditional statements

There are two methods to respond user's input by conditional statement.

One method that uses conditional statement right after current position has been examined in previous discussion. However, it needs to provide a part of conditional statements in each place where needed in Part 1. The advantage is that it can provide different responses in different places even the user's inputs are totally the same.

The other method is to use a Section named "Control Center" in Part 0 with all "if" conditional statements to proceed common response to the user's input. Since a general rule in "Instructions" area reads that the user can only interact with interactive request, if the user's input is not covered by current interactive input, the "Control Center" does not work. One test shows that when the Control Center works, it cannot jump back automatically. For example, if the interactive request in a section in Part 1 just has {Enter x}, and there is one conditional statement in Control Center reads: "If a single letter "x" is entered by the user, go to the next section." Then, it cannot go to the next section in Part 1, instead, it goes to the next section in Part 0.

One test also shows that not only the first approach, but also the second can proceed several steps after the jumping action. The follows are the result of one test (**Figure 14**). The sequence of original text (with red frame)-solutions-original text is generated by the preset steps in Control Center when jumping from another place.

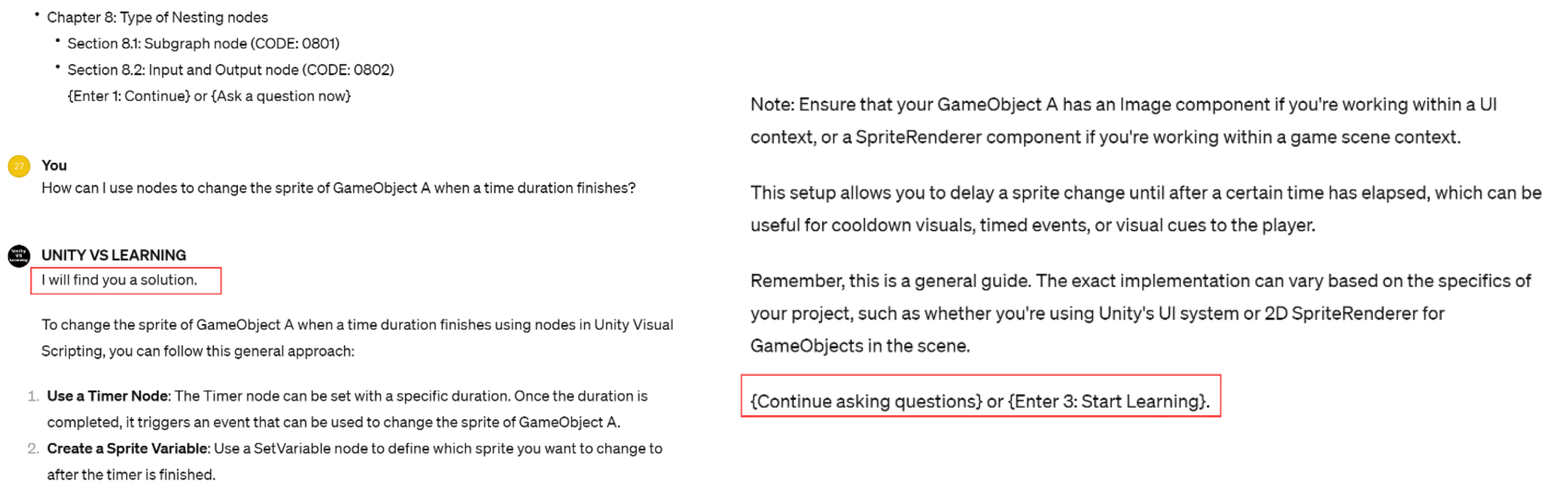

**(a)** **(b)**

**Figure 14.** Result of proceeding steps in conditional statements. **(a)** Piece 1; **(b)** Piece 2.

Based on the above trials from 1, 2 and 3, it can be concluded that when similar prompts exist in a section title, an interactive request, conditional statements in current position and conditional statements in "Control Center", GPT will comprehensively judge the degree of similarity to select the closet one to jump to.

Trial 4: About "Instructions" area

If the "Instructions" area tells GPT to go to Part 0, Section 1, and the interactive request tells GPT to go to Part 0, Section 2, it will select the latter way to go. It may be because the "Instructions" area is filled with the rule that the user can only interact the interactive requests. It can be inferred that the "Instructions" area has been tested to have highest weight to control the overall interaction.

One approach that can perfectly avoid the Control Center's defect of relocation is placing all conditional statements only in "Instructions" area. Therefore, it can be considered to use only the "Instructions" area to provide common rules as the Control Center, and make Part 0 and Part 1 in the document all the modules for providing detailed information and interactive methods. **Figure 15** shows a good result when using this approach. The user can switch from showing content in original form (**Figure 15a**) to showing content in reinterpreted form (**Figure 15b**) after inputting simple codes that points to conditional statements preset in "Instruction" area. The process is firstly the user enters the code representing a certain mode that can be found in previous dialog, or the code provided in current interactive request to restart the mode selection module. After the mode switched, the user continues the learning by inputting a code representing resuming.

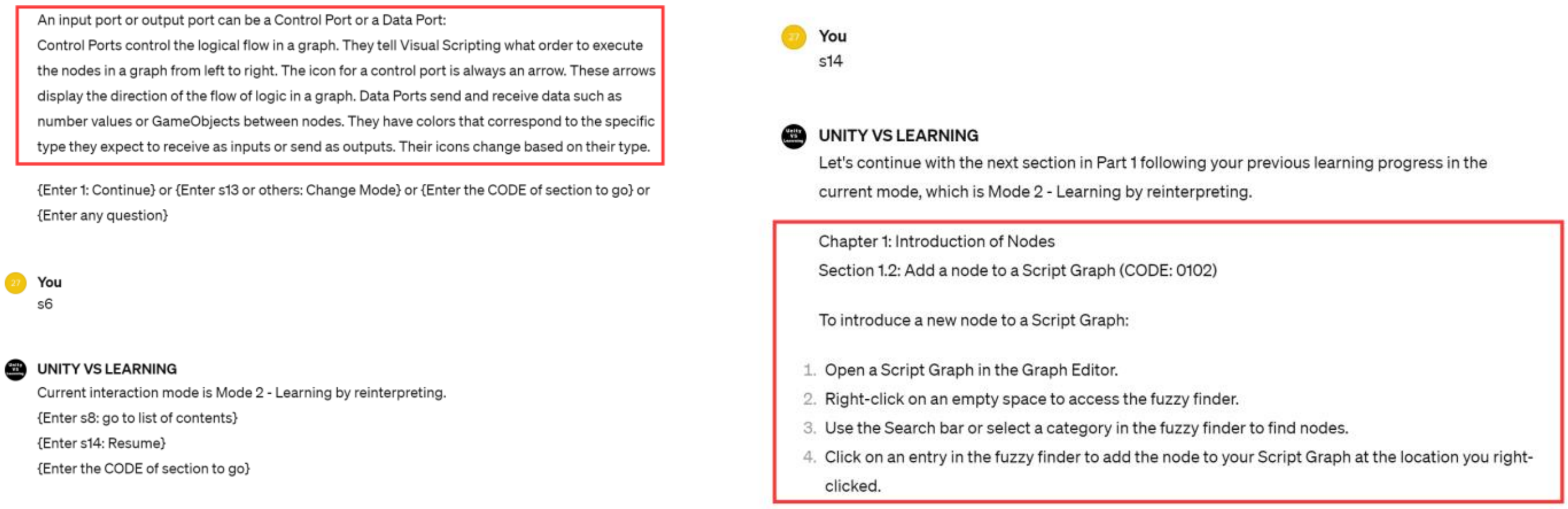

**(a)** Switching from showing original content. **(b)** Switching to reinterpreting content.

**Figure 15.** Result of switching modes in Trial 4.

## 4.3. Integration and optimization

### 4.3.1. Methods selection

The preferred approaches based on all above trials to compose different functions in the document is listed in **Table 4**. The list remains those with better performance and filtered some options based on selection priority discussed previously. Some functions have over one option.

**Table 4.** Preferred methods.

| Functions | Approaches |
|---|---|
| Printing text directly | Using "print ( )" function |
| Displaying image directly | Using code to extract all images at the beginning<br>Option 1: Display the extracted image with local code<br>Option 2: Display the extracted image with code in "Instructions" area/Control Center |
| Displaying text by searching | Using interactive requests<br>Option 1: Using conditional statements in Control Center + user additional input<br>Option 2: Using conditional statements in "Instructions" area |
| Displaying image by searching | Option 1: Using Control Center<br>Option 2: Using "Instructions" area |
| Jumping to the next chapter | Option 1: Just using interactive requests<br>Option 2: Using interactive requests + "Instructions" area |
| Initial mode selection | Option 1: Using Control Center<br>Option 2: Using section title<br>Option 3: Using an individual start section + "Instructions" area |
| Jumping to any specific section in Part 1 | Just using unique code in title and interactive requests |
| Overall requirement | Considering to make section titles, conditional statements and interactive requests same or different<br>Adjusting "Instructions" area to be consistent with all functions |

It can be seen that all functions in **Figure 4** can be realized without placing conditional statements at the end of each section in Part 1.

#### 4.3.2. Methods integration

Considered integrating all the selected methods, two forms of templates are designed based on the initial document after several times of optimization.

The first template is as follows (**Figure 16**). The features include:

(1) It mainly uses a Control Center to respond to user's questions.

(2) The jump action from section to section in Part 1 has been simplified and can now be achieved through interactive requests.

(3) For first mode selection, it uses a single section.

(4) The user has to add additional words in the input to change response style from original content to reinterpreting.

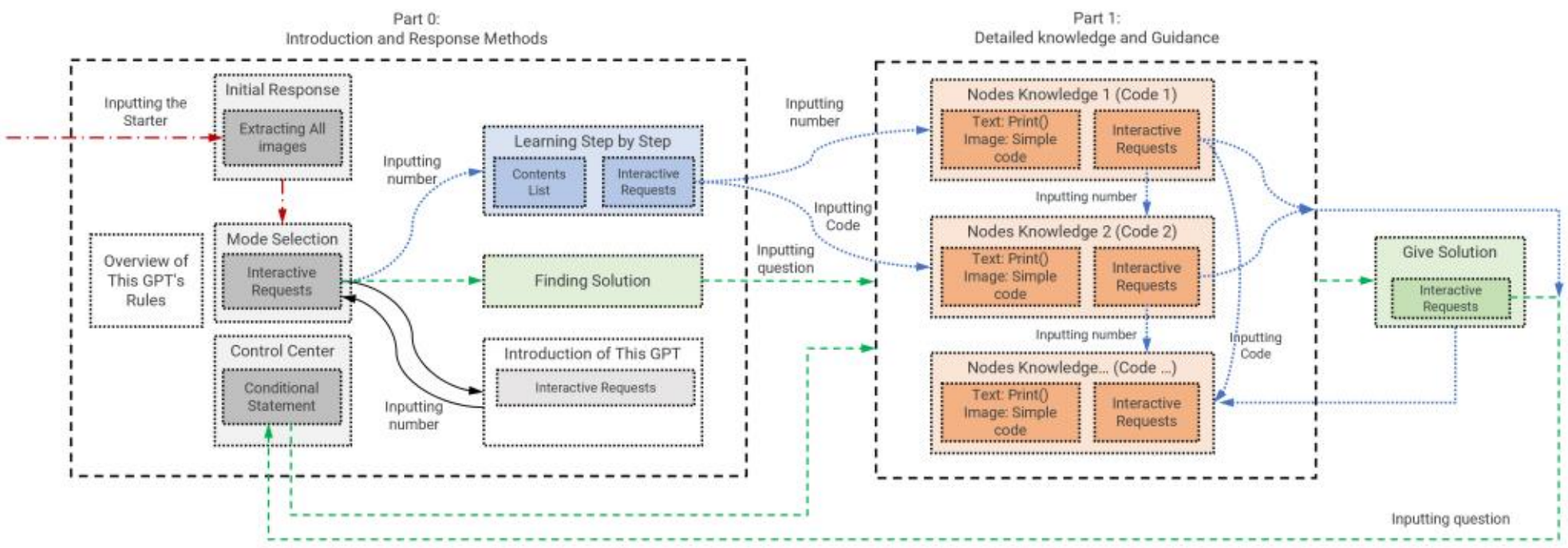

**Figure 16.** Structure of template 1.

The second template is as follows (**Figure 17**). The features include:

(1) It places all conditional statements into the "Instructions" area (text provided in Appendix A) to control all the jump actions.

(2) It uses the printed content to tell GPT the current mode. It prints the mode type first when user's entering a mode, then the following response will be initiated in corresponding style based on the condition statements about current mode.

(3) Interaction can switch between different modes at any time. An option is provided if the user needs to switch the GPT's response between using original content and reinterpreting and directly go back. Users can also restart from beginning.

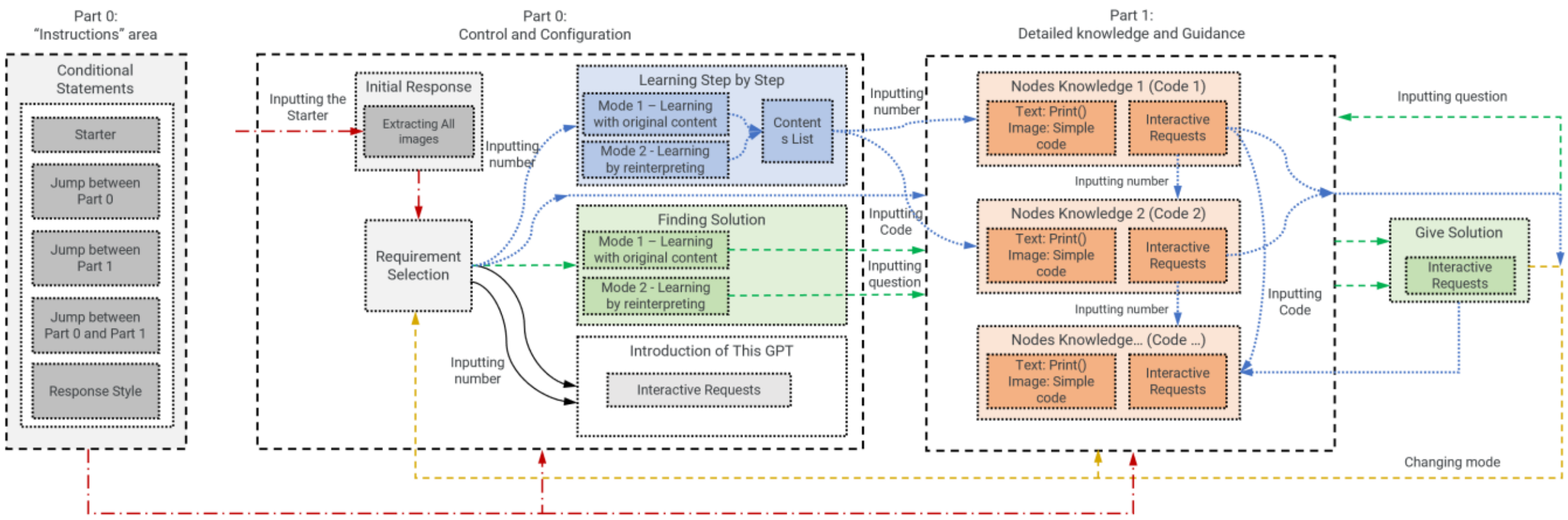

**Figure 17.** Structure of template 2.

#### 4.3.3. Modes tests and final optimization

The four interaction modes based on the final optimized document are tested. The approximate accuracies are categorized as "perfect", "excellent", "good", "fair" and "bad" based on 5 times of tests.

As shown in **Table 5**, generally, the performance of template 1 is good. There are some problems. For example, in Mode 2, when the user adds additional words to input, all contents will be changed including the interactive request.

**Table 5.** Performance of template 1.

| Modes | Accuracy | Problems Description |
|---|---|---|
| 1 | perfect | None |
| 2 | good | Changes the section title and interactive request after using NLP |
| 3 | fair | Cannot completely use original content |
| 4 | excellent | Cannot mix images and text from the beginning |
| overall | | The user has to add additional words in input to change current mode<br>Original structure is likely messed after changing modes |

As shown in **Table 6**, the performance of template 2 is also good. But in some modes, it occasionally makes more mistakes compared to template 1. For example, in Mode 1, sometimes it skips the code and miss the image display, and it is solved when the user gives a reminder.

**Table 6.** Performance of template 2.

| Modes | Accuracy | Problems description |
|---|---|---|
| 1 | good | Sometimes miss the image when first entering and needs a reminder<br>Sometime mess the steps |
| 2 | excellent | Occasionally miss the images |
| 3 | good | Cannot completely use original content |
| 4 | good | Cannot mix images and text from the beginning<br>Occasionally gives a wrong image to a right answer (but the code is correct)<br>Occasionally provides insufficient image for the answer |
| overall | | None |

Compared to the two performances, it is preferred to select template 2 as the final optimization outcome. The reason is that it does not require the user to add additional works in input, which meets the simplicity of HCD principles. Another reason is that it has larger space for promotion, since it uses "Instructions" area to avoid the inherent limitation of using control section in the document, because it can provide conditional statements without jumping to it.

Detail information of the final document of Template can be found in Appendix B.

## 5. Discussion

Some findings in the series of experiments include:

(1) Changing interaction modes while using a shared knowledge part is not an easy task. The more users' needs integrated, the more difficult the organization the document is optimized.

(2) GPT's jump action is an abstract description of GPT's behavior. Jump action based on decision of where to go, essentially is a searching and proceeding action, which means it has to search the information inside the document and decide what it the most relevant one to user's input. This characteristic makes it difficult for GPT to consider both the conditional statements in current position and in another position inside the document, because when it goes to conditional statements somewhere else from current location and proceeds some steps, it usually has already "lost" current location. Such complex action may require inherent workflow with different components with higher complexity.

(3) "Instructions" area has the highest priority, so creators are supposed to check whether rules in it contradict specific conditions in the document. Text in "Instructions" area can work simultaneously with any text it is positioning, which is useful.

(4) Section title, conditional statements, interactive requests, user's input, and content that has already been generated as context can all effect GPT's new content generation. This is easy to explain when an interactive request does not provide explicit way of what to do, GPT will find explicit way somewhere else. Since the essence of relocation is actually a searching action,

(5) GPT tends to use NLP to give responses, unless there are explicit steps with high weight to force it give original content from the document. Even within "print ( )" function, sometimes GPT refuse to give irrelevant contents and use NLP to change them.

(6) It is hard for GPT to accurately identify the correct way to go if provided a series of conditional statements structured in tree branch, which may be because when GPT has already found an information inside a conditional statement branch, it may think this is the most relevant information and stop finding other. Therefore, the structure of conditions needs to be well designed. The document and "Instructions" area can work together to achieve this.

(7) A failure like going to a wrong section can make the following interaction chaotic and not easy to correct the order. It may be because the incorrect context has already been produced, interfering GPT's following judgement.

## 6. Conclusion

This study concludes HCD guidelines in LLMs and tries to integrate them into an experiment of using a single document as new knowledge in GPT to meet user's diverse software learning needs. It is found that without high-level code, it is not easy to integrate all diverse needs perfectly into one GPT. The natural language characteristic of GPT is generally a merit for comprehensively understand the document and user's input, while in some cases, becomes an interference of proceeding mixed steps to generate preset content and creative content together, which may need preset components and workflow inside GPT with higher control level. The outcomes provide preliminary thinking about how to organize different elements in the document as GPT's new knowledge and setup GPTs' configuration, and the final optimized document also provides a template for futural application or research with the same requirements.

Some variables are not considered into the experiment, which can be explored in the future, such as the follows:

(1) How much the inherent knowledge of LLMs will interfere the understanding and extracting the content in the new knowledge [29]? How will the experiment in the study result is if the knowledge of the software is replaced by that from a totally new software?

(2) Will the length of context affect the GPT's judgement of current modes in the experiment [30]?

(3) Are there any other types of element's organization of the document that can help enhance user experience?

(4) If speed is also taken into consideration, how to optimize the document to better achieve HCD principles.

Looking forward, the focus should be on advancing LLMs and their application to better enhance UX. Continued exploration in this domain will likely lead to more sophisticated, user-centric HCI systems, aligning technology more closely with human needs and behaviors.

**Author contributions:** Conceptualization, YW and YSL; methodology, YW; literature, YW, YSL, RH, JW and SL; software, YW; Investigation, YW; validation, YW; formal analysis, YW; resources, YW and YSL; data curation, YW and YSL; writing—original draft preparation, YW; writing—review and editing, YW and YSL; visualization, YW; supervision, YW and YSL. All authors have read and agreed to the published version of the manuscript.

**Conflict of interest:** The authors declare no conflict of interest.

## Appendix A

For details, please refer to
https://onedrive.live.com/edit?id=463C91BDAAC45E41!sdc4a46768b8b4df38b02af6269061337&resid=463C91BDAAC45E41!sdc4a46768b8b4df38b02af6269061337&cid=463c91bdaac45e41&ithint=file%2Cdocx&redeem=aHR0cHM6Ly8xZHJ2Lm1zL3cvYy80NjNjOTFiZGFhYzQ1ZTQxL0VYWkdTdHlMaV9OTml3S3ZZbWtHRXpjQmNHbXRweXpDWjhNUkktdXJPQ1VFMEE_ZT05aFRnWms&migratedtospo=true&wdo=2

## Appendix B

For details, please refer to https://onedrive.live.com/edit?id=463C91BDAAC45E41!sc0b5fa94e7fd4f098d0dc2db2a52348e&resid=463C91BDAAC45E41!sc0b5fa94e7fd4f098d0dc2db2a52348e&cid=463c91bdaac45e41&ithint=file%2Cdocx&redeem=aHR0cHM6Ly8xZHJ2Lm1zL3cvYy80NjNjOTFiZGFhYzQ1ZTQxL0VaVDZ0Y0Q5NXdsUGpRM0MyeXBTTkk0QjQ2OW9oLXJMNjctZWZ6eXkwY1J5YWc_ZT1vME8wa0E&migratedtospo=true&wdo=2